\documentclass[aps,preprint,epsfig,rotate]{revtex4}
\begin{document}
\title{The hyperfine structure of the ground states in the helium muonic atoms}

 \author{Alexei M. Frolov}
 \email[E--mail address: ]{afrolov@uwo.ca}

\affiliation{Department of Chemistry\\
 University of Western Ontario, London, Ontario N6H 5B7, Canada}

\date{\today}

\begin{abstract}

The hyperfine structures of the ground states in the ${}^{3}$He$^{2+} \mu^{-} 
e^{-}$ and ${}^{4}$He$^{2+} \mu^{-} e^{-}$ helium-muonic atoms are 
investigated with the use of highly accurate variational wave functions. The 
differences between corresponding levels of hyperfine structure (i.e. hyperfine
splittings) are determined to very high numerical accuracy. In particular, we 
have found that the hyperfine structure splitting in the ground state of the 
${}^{4}$He$^{2+} \mu^{-} e^{-}$ atom is $\Delta \nu \approx 4464.55454(3)$ 
$MHz$, while analogous splitting for the ${}^{3}$He$^{2+}\mu^{-}e^{-}$ atom is
$\Delta \nu \approx 4166.43547(3)$.

\end{abstract}

\maketitle
\newpage

\section{Introduction}

In this study we report the results of highly accurate calculations of the 
ground $S(L = 0)-$states in the helium-muonic ${}^{3}$He$^{2+} \mu^{-} e^{-}$
and ${}^{4}$He$^{2+} \mu^{-} e^{-}$ atoms. Our recently improved methods for
highly accurate variational computations allow us to construct extremely 
accurate variational wave functions for these three-body helium-muonic atoms.
Such wave functions can be used to obtain essentially exact expectation 
values of various bound state properties of these systems. In particular, 
the expectation values of all interparticle delta-functions $\delta_{ij}$ 
can now be determined to very high numerical accuracy. By using these 
expectation values one can finally solve the long standing problem of 
hyperfine structure splitting in the ${}^{3}$He$^{2+} \mu^{-} e^{-}$ and 
${}^{4}$He$^{2+} \mu^{-} e^{-}$ atoms \cite{Lak1} - \cite{Fro02}. 

In this study we reconsider the problem of highly accurate computations of
the helium-muonic ${}^{3}$He$^{2+} \mu^{-} e^{-}$ and ${}^{4}$He$^{2+} \mu^{-} 
e^{-}$ atoms. For simplicity, below we restrict ourselves to the analysis of 
the ground $S(L = 0)-$states in such systems. The ground state in the 
helium-muonic atom is designated as the $1s_{\mu}1s_e$-state. Analogous 
electron-excited states (or $1s_{\mu}2s_e$-state) in the helium-muonic atoms
were discussed in our earlier study \cite{Fro02}. Our main goal is to determine
the hyperfine structure splitting in each of these atoms by using the results 
of most recent highly accurate computation of the expectation values of  
interparticle delta-functions. The computed values of the hyperfine structure 
splitting in each helium muonic atom must be compared with the known 
experimental results \cite{muon}. For the ground states in the helium-muonic 
atoms such experiments were performed in 1970's. Analogous experimental 
observations and measurments of the hyperfine structure splitting for the 
electron-excited states in the helium-muonic atoms are difficult to perform. 

For the ground states in the ${}^{3}$He$^{2+} \mu^{-} e^{-}$ and ${}^{4}$He$^{2+} 
\mu^{-} e^{-}$ helium-muonic atoms the hyperfine structure spllittings have been 
measured to very good accuracy (see references and discussions in \cite{muon}). 
The known experimental values of the hyperfine structure splitting of the ground 
states in these two atoms are 4166.41 MHz and 4464.95 MHz, respectively 
\cite{muon}. A large number of theoretical investigations of muonic-helium atoms 
have been conducted since the middle of 1970's. In many of these works the main 
goal was to produce highly accurate evaluation of the hyperfine structure splitting 
for the ground states of the ${}^{3}$He$^{2+} \mu^{-} e^{-}$ and ${}^{4}$He$^{2+} 
\mu^{-} e^{-}$ atoms. In general, to obtain the actual hyperfine structure and 
evaluate the hyperfine structure splittings one needs to detemine the expecation 
values of all three interparticle delta-functions. The main computational troubles 
of earlier studies were related with the expectation values of the electron-muonic 
delta-function. Another complication was related to a slow convergence rate 
observed in numerical computations of the expectation values of the electron-nucleus 
delta-function. In general, it was very hard to determine the electron-muonic and 
electron-nucleus delta-functions to the accuracy better than $1 \cdot 10^{-6}$ 
$a.u.$ For the electron-nucleus delta-function such a `limiting' accracy was 
$\approx$ $2 \cdot 10^{-7}$ $a.u.$ Finally, it was almost impossible to evaluate the 
hyperfine structure splitting in the helium-muonic atoms to good numerical accuracy. 
A substantial progress has been achieved only in our work \cite{Fro02} where we have 
applied a special Fortran pre-translator MPFUN written by David H.Bailey \cite{Bail1}, 
\cite{Bail2}. By using this pre-translator we determined the hyperfine structure 
splitting 4166.392 $MHz$ for the ground state in the ${}^{3}$He$^{2+} \mu^{-} 
e^{-}$ atom and 4464.555 $MHz$ for the ground state in the ${}^{4}$He$^{2+} \mu^{-} 
e^{-}$ atom \cite{Fro02}. The absolute uncertainties in both cases are less than 10 
kHz. It appears that these values were very close to the known experimental values 
of the hyperfine structure splitting measured for these atoms, which equal 4166.41 
$MHz$ and 4464.95 $MHz$, respectively \cite{muon}.

In this study we substantially improved the overall accuracy of our computations. The
overall quality of our wave functions has also been improved. The current accuracy of 
the expectation values of interparticle delta-functions can now be evaluated as $\pm 7 
\cdot 10^{-11}$ $a.u.$ (or even better).  Right now, we can determine the hyperfine 
structure splittings in both helium muonic atoms to much better accuracy than it was 
possible in \cite{Fro02}.  

\section{Three-body variational wave function}

Our computational goal in this study is to determine the highly accurate solutions of the 
non-relativistic Schr\"{o}dinger equation $H \Psi({\bf r}_e, {\bf r}_{\mu}, {\bf r}_{\rm 
He}) = E \Psi({\bf r}_e, {\bf r}_{\mu}, {\bf r}_{\rm He})$, where $E < 0$ is the numerical 
parameter (energy) and $H$ is the following Hamiltonian
\begin{eqnarray} 
  H = -\frac{\hbar^2}{2 m_e} \Bigl[ \nabla_e^2 + \Bigl(\frac{m_e}{m_{\mu}}\Bigr) 
 \nabla_{\mu}^2 + \Bigl(\frac{m_e}{M_{\rm He}}\Bigr) \nabla_{\rm He}^2 \Bigr] +
 \frac{e^2}{r_{e \mu}} - \frac{2 e^2}{r_{e {\rm He}}} - \frac{2 e^2}{r_{\mu {\rm He}}}
 \label{Hamil}
\end{eqnarray}
where $\hbar, m_e$ and $e$ are the reduced Planck constant, electron mass and absolute 
value of the electric charge of the electron. Everywhere below in this study, we shall 
use only atomic units, where $\hbar = 1, m_e = 1$ and $e = 1$. In these units one finds
\begin{eqnarray} 
  H = -\frac12 \Bigl[ \nabla_e^2 + \Bigl(\frac{1}{m_{\mu}}\Bigr) 
 \nabla_{\mu}^2 + \Bigl(\frac{1}{M_{\rm He}}\Bigr) \nabla_{\rm He}^2 \Bigr] +
 \frac{1}{r_{e \mu}} - \frac{2}{r_{e {\rm He}}} - \frac{2}{r_{\mu {\rm He}}}
 \label{Hamilau}
\end{eqnarray}
where masses of the muon $m_{\mu}$ = 206.768262 $m_e$ and helium nuclei $M_{{}^3{\rm He}}$ 
= 5495.8852 $m_e$ and $M_{{}^4{\rm He}}$ = 7294.2996 $m_e$ must be expressed in the 
electron mass $m_e$.  

To simplify all formulas below we designate the electron, muon and helium nucleus as 
particles 1, 2 and 3, respectively. For the considered ground $S(L = 0)-$states in the 
three-body ${}^{3}$He$^{2+} \mu^{-} e^{-}$ and ${}^{4}$He$^{2+} \mu^{-} e^{-}$ muonic 
atoms the wave function depends upon three scalar interparticle coordinates, i.e.      
$\Psi({\bf r}_e, {\bf r}_{\mu}, {\bf r}_{\rm He}) = \Psi(r_{21}, r_{31}, r_{32})$. 
This unknown wave function $\Psi$ can be approximated to very high accuracy by using 
different variational expansions. In particular, in this work we apply the exponential 
variational expansion in three-body relative coordinates $r_{21}, r_{31}$ and $r_{32}$. 
For the $S(L = 0)-$states in the helium-muonic atoms this expansion takes the form
\begin{eqnarray}
 \Psi_{LM} = \sum_{i=1}^{N} C_{i} \exp(-\alpha_{i} r_{21} -\beta_{i} r_{31} - 
 \gamma_{i} r_{32}) \label{exp1}
\end{eqnarray}
where $C_{i}$ are the linear (or variational) parameters, while $\alpha_i, \beta_i, 
\gamma_i$ are the non-linear parameters. In general, the exponential variational expansion
has significantly more complex form even for the $S(L = 0)-$states \cite{Fro06}. First, it is 
convenient to write such an expansion in the three perimetric coordinates $u_1, u_2, u_3$ 
which are completely independent and each of them varies between 0 and $+\infty$. In this 
case all non-linear parameters $\alpha_i, \beta_i, \gamma_i$ must be positive. Second, for 
some three-body systems it is necessary to use the complex values for these parameters. It
is allows one to accelerate the overall convergence rate of this variational expansion, e.g.,
for adiabatic and/or weakly-bound systems. Here we do not want to discuss all advanteges of
exponential variational expansions for three-body systems. Note only that the expansion,
Eq.(\ref{exp1}), provides a simple method \cite{Fro06} to construct extremely accurate 
variational wave functions for an arbitrary three-body systems (and different states in such 
systems). The overall quality of the final wave function, Eq.(\ref{exp1}) is very good, and it 
can be used to determine all expectation values needed for analysis of the given three-body 
system. In particular, such a wave function generates very accurate expectation values of all 
interparticle delta-functions which are of great interest for this study (see the next 
Section). 

\section{Spin-spin contact interaction between two particles}

Let us assume that we have two particles $a$ and $b$ which have non-zero spins 
${\bf s}_a$ and ${\bf s}_b$, respectively. The non-relativistic Hamiltonian of 
the spin-spin interaction takes the form 
\begin{eqnarray} 
  H_{ss} = - \frac{\mu_a}{s_a} {\bf s}_a \cdot {\bf H}_b = - \frac{\mu_a}{s_a} 
  \frac{1}{c} \Bigl({\bf s}_a \cdot \int \frac{{\bf n} \times {\bf j}_b}{r^2} dV 
  \Bigr)
\end{eqnarray}
where $c$ is the speed of light in vacuum, ${\bf n} = \frac{{\bf r}}{r}$ is the 
unit vector which corresponds to the radius-vector ${\bf r}$ and ${\bf j}_b$ is
the current generated by the particle $b$ with spin ${\bf s}_b$. For this current 
we can write 
\begin{eqnarray} 
  {\bf j}_b = \frac{\mu_b}{s_b} c \; (\nabla \times (\Psi^{*} {\bf s}_b \Psi)) = 
  \frac{\mu_b}{s_b} c \; \Bigl(\frac{d \Psi^2(r)}{dr} \Bigr) ({\bf n} \times 
  {\bf s}_b)
\end{eqnarray}
The magnetic field ${\bf H}_b$ generated by this spin current ${\bf j}_b$ takes the 
form 
\begin{eqnarray} 
 {\bf H}_b = \frac{\mu_b}{s_b} \int_{0}^{\infty} \frac{d \Psi^2(r)}{dr} dr
 \oint ({\bf n} \times ({\bf n} \times {\bf s}_b)) do = 
 \frac{\mu_b}{s_b} \Psi^2(0) \frac{8 \pi}{3} {\bf s}_b
\end{eqnarray}
Therefore, the Hamiltonian of the spin-spin interaction of the particles $a$ and $b$ is 
\begin{eqnarray} 
  H_{ss} = -\frac{8 \pi}{3} \frac{\mu_a \mu_b}{s_a s_b} ({\bf s}_a \cdot 
 {\bf s}_b) \langle \delta({\bf r}_{ab}) \rangle  \label{e3}
\end{eqnarray}
where $\delta({\bf r}_{ab}) = \delta({\bf r}_a - {\bf r}_b)$ is the delta-function
between two particles ($a$ and $b$). The interparticle interaction which contains
the corresponding delta-function is called the contact interaction.

In atomic systems all magnetic moments are traditionally measured in Bohr magnetons
$\mu_B$. In atomic units $\hbar = 1, m_e = 1, e = 1$ the value of the Bohr magneton
equals $\frac12$ exactly. In these units the formula Eq.(\ref{e3}) takes the form 
\begin{eqnarray} 
  H_{ss} = -\frac{2 \pi}{3} \frac{\mu_a \mu_b}{s_a s_b} \langle \delta({\bf r}_{ab}) 
 \rangle ({\bf s}_a \cdot {\bf s}_b) \label{e35}
\end{eqnarray}
In a few-body system the total Hamiltonian of the spin-spin interaction equals to the
sum of different two-particle Hamiltonians, Eq.(\ref{e35}). In the general case, in 
$A-$body system one finds $\frac{A (A-1)}{2}$ spin-spin terms, Eq.(\ref{e35}), in the 
total Hamiltonian which describes the hyperfine interactions. In reality, many actual 
few-body systems contain identical particles and, therefore, the total number of terms 
in the Hamiltonian of spin-spin interaction is less than the number of interparticle 
interactions $\frac{A (A-1)}{2}$.

\section{The hypefine structure of the helium-muonic atoms}

It follows from here that the general formula for the Hamiltonian of hyperfine 
structure $(\Delta H)_{h.s.}$ of an arbitrary three-body system is written as 
the sum of the three following terms. Each of these terms is proportional to the 
product of the factor $\frac{2 \pi}{3} \alpha^2$ and expectation value of the 
corresponding (interparticle) delta-funtion. The third (additional) factor 
contains the corresponding $g-$factors (or hyromagnetic ratios) and scalar 
product of the two spin vectors. For instance, for the ${}^{3}$He$^{2+} \mu^{-} 
e^{-}$ atom this formula takes the form (in atomic units) (see, e.g., \cite{LLQ}, 
(see also \cite{Fro02}))
\begin{eqnarray}
 (\Delta H)_{h.s.} = \frac{2 \pi}{3} \alpha^2 \frac{g_{N} g_{\mu}}{m_p m_{\mu}}
 \langle \delta({\bf r}_{32}) \rangle ({\bf I}_{N} \cdot {\bf s}_{\mu})+
 \frac{2 \pi}{3} \alpha^2 \frac{g_{N} g_{e}}{m_p m_{e}}
  \langle \delta({\bf r}_{31}) \rangle ({\bf I}_{N} \cdot {\bf s}_{e}) 
 \nonumber \\
 + \frac{2 \pi}{3} \alpha^2 \frac{g_e g_{\mu}}{m_e m_{\mu}}
  \langle \delta({\bf r}_{21}) \rangle ({\bf s}_e \cdot {\bf s}_{\mu})
 \label{hspl1}
\end{eqnarray}
where $\alpha = \frac{e^2}{\hbar c}$ is the fine structure constant, $m_{\mu}$ 
and $m_p$ are the muon and proton masses, respectively. The notation $N$ stands
for the nucleus ${}^3$He. The factors $g_{\mu}, g_{e}$ and $g_{N}$ are the 
corresponding $g-$factors. The expression for $(\Delta H)_{h.s.}$ is, in fact, 
an operator in the total spin space which has the dimension $(2 s_{N} + 1) 
(2 s_{\mu} + 1) (2 s_{e} + 1) = 8$. In our calculations we have used the 
following numerical values for the constants and factors from Eq.(\ref{hspl1}): 
$\alpha = 7.2973525698\cdot 10^{-3}, m_p = 1836.152701 m_e, m_{\mu} = 206.768262 
m_e$, $g_{\mu}$ = -2.0023318414 and $g_e$ = -2.0023193043622. The $g-$factor of 
the helium-3 nucleus is deteremined from the formula: $g_{N} = \frac{{\cal 
M}_{N}}{I_{N}}$ = -4.2555016, where ${\cal M}_{N}$ = -2.1277508 \cite{CRC} is the 
magnetic moments (in nuclear magnetons) of the helium-3 nucleus. The spin of the 
helium-3 nucleus is designated in Eq.(\ref{hspl1}) as $I_{N} = \frac12$. 

This means that in the ground state of the helium-3 muonic atom one finds eight 
spin states. They are separated into three different groups: (1) four spin states 
with $J = \frac32$, (2) two `upper' spin states with $J = \frac12$, and (3) two 
`lower' spin states with $J = \frac12$. Here and below the notation $J$ (or ${\bf 
J})$ stand for the total angular momentum ${\bf J} = {\bf L} + {\bf S}$. For the 
$S(L = 0)-$states we have ${\bf J} = {\bf S}$, and therefore, $J = S$. The energy 
of the group of four states with $J = \frac32$ is $\approx$ 8.296777691701$\cdot 
10^7$ $MHz$, while the energy of the second group of two `upper' spin states with 
$J = \frac12$ is $\approx$ 8.296361048154$\cdot 10^7$ $MHz$. The group of two 
`lower' spin states with $J = \frac12$ has the energy $\approx -2.488991643156\cdot 
10^8$ $MHz$. The energy difference between the first and second groups of spin 
states is $\approx 4166.43547(3)$ $MHz$. Such a difference is called the hyperfine 
structure splitting of the ground state of the helium-3 muonic atom. To recalculate 
the energies expressed in atomic units into $MHz$ we have used the following 
conversion factor $Ry$ = 6.579683920729$\cdot 10^9$.

The spin and magnetic moment of the helium-4 nucleus equals zero identically.
Therefore, the hyperfine structure splitting in the ground state of the 
${}^{4}$He$^{2+} \mu^{-} e^{-}$ atom is related with the electron-muon spin-spin
interaction. Four spin state of the electron-muon pair are separated in the two
groups with $J = 1$ (three states) and $J = 0$ (one state). The energy difference
between these two groups of states is called the hyperfine structure splitting of 
the ground state in the ${}^{4}$He$^{2+} \mu^{-} e^{-}$ atom. With the constants 
mentioned above this hyperfine structure splitting is determined with the use of 
the formula
\begin{eqnarray}
 \Delta \nu = 14229.178083766834 \cdot \langle \delta({\bf r}_{e^{-} \mu^{-}}) 
 \rangle
\end{eqnarray}
where $\langle \delta({\bf r}_{e^{-} \mu^{-}}) \rangle$ is the expectation value 
of the electron-muon delta-function. This formula produces the hyperfine structure 
splittings $\Delta \nu$ in $MHz$, if the expectation value of delta-functions is 
given in atomic units. By using this formula and our expectation value of the 
electron-muon delta-function from Table II we have found that $\Delta 
\nu({}^{4}$He$^{2+} \mu^{-} e^{-})$ = 4464.55454(3) $MHz$. 

The uncertanties mentioned above have been determined from the corresponding 
uncertainties of the delta-functions. For instance, for the ${}^{4}$He$^{2+} \mu^{-} 
e^{-}$ atom one finds (from Table II) for the $\langle \delta({\bf r}_{e^{-} 
\mu^{-}}) \rangle$ expectation value $\langle \delta({\bf r}_{e^{-} \mu^{-}}) 
\rangle \approx$ 3.13760535832(7)$\cdot 10^{-1}$ $a.u.$ Now, by using three 
expectation values: 3.13760535825$\cdot 10^{-1}$, 3.13760535832$\cdot 10^{-1}$ and
3.13760535839$\cdot 10^{-1}$ we have evaluated the corresponding uncertanty in the
$\Delta \nu({}^{4}$He$^{2+} \mu^{-} e^{-})$ value as $\approx 1 \cdot 10^{-5}$ $MHz$.
By including the uncertainties known from modern experiments for the fine structure
constant $\alpha$, conversion factor $Ry$, masses of the particles, etc one finds 
that our maximal uncertanty can be evaluated as $\approx 3 \cdot 10^{-5}$ $MHz$.

It should be mentioned that our hyperfine structure splittings $\Delta \nu$ 
determined to very high accuracy in this study are very sensitive to variations
of the basic physical constants, e.g., masses of particles, fine structure constant
$\alpha$, etc. In the case of the fine-structure constant $\alpha$ the actual 
changes of the numerical values of $\Delta \nu$ are quadratic upon $\alpha$. This
fact can be used to develop new experimental methods to measure the fine structure
constant $\alpha$ and particle masses $m_{\mu}, M_{{}^3{\rm He}}, M_{{}^4{\rm He}}$ 
to very high accuracy.    

\section{Conclusion}

We have considered the hyperfine structures of the ground states of the helium-3 
and helium-4 muonic atoms. Our predicted values of the hyperfine structure 
splittings for the helium-3 and helium-4 muonic atoms are 4166.43547(3) 
$MHz$ and 4464.55454(3) $MHz$, respectively. They are very close to the known 
experimental values 4166.41 MHz and 4464.95 MHz, respectively \cite{muon}. Note that the 
corresponding experiments have been performed in 1970's. Since then no new accurate 
measurments of the hyperfine splittings in these muonic atoms have been reported. In 
reality, such experiments are desperately needed to explain currently observed 
deviations between highly accurate theoretical and experimental results. For instance, 
the difference between theoretical and experimental results for the helium-3 muonic 
atom is $\approx$ 20 times smaller than for the helium-4 atom. This has never been
expalined. On the other hand, the new experiments are also needed to predict and 
explain some new phenomena, e.g., to determine hyperfine structure splittings in the 
electron excited states of the helium-muonic atoms. Their predicted values can be found
in \cite{Fro02}.

\newpage
\begin{table}[tbp]
\caption{The convergence of the total energies in atomic units for
         the ground $1s_{\mu}1s_e-$states in the helium-muonic atoms.
         $N$ is the total number of basis functions used in 
         calculations.}
  \begin{center}
    \begin{tabular}{lll}
      \hline\hline
$N^{a}$ & ${}^3$He$^{2+}\mu^{-}e^{-}$ & ${}^4$He$^{2+}\mu^{-}e^{-}$ \\
     \hline
3300      & -399.042 336 832 862 534 826 9920 & -402.637 263 035 135 454 018 9238 \\

3500      & -399.042 336 832 862 534 826 9993 & -402.637 263 035 135 454 018 9313 \\

3700      & -399.042 336 832 862 534 827 0052 & -402.637 263 035 135 454 018 9373 \\

3840      & -399.042 336 832 862 534 827 0088 & -402.637 263 035 135 454 018 9410 \\
        \hline
$P^{(a)}$ & -399.042 336 832 862 534 769     & -402.637 263 035 135 454 004 81 \\
 \hline\hline
 \end{tabular}
 \end{center}
${}^{(a)}$The best results determined in earlier calculations [8].
\end{table}
\begin{table}[h]
\caption{Convergence of the expectation values of electron-muon and
         electron-nucleus delta-functions in atomic units for the ground
         $1s_{\mu}1s_e-$states in the ${}^{3}$He$^{2+}\mu^{-}e^{-}$ and
         ${}^{4}$He$^{2+}\mu^{-}e^{-}$ helium-muonic atoms. The notation 1
         means the electron $e^{-}$, the notation 2 stands for the muon 
         $\mu^{-}$ and notation 3 designates the nuclues of the helium-3 
         and helium-4, respectively.}
  \begin{center}
   \begin{tabular}{ccccc}
       \hline\hline
$N({}^3$He) & $\langle \delta({\bf r}_{21}) \rangle$ & $\langle 
\delta({\bf r}_{31}) \rangle$ & $\langle \delta({\bf r}_{32}) \rangle$ & 
$\langle \delta({\bf r}_{321}) \rangle$ \\
      \hline
3300 & 3.13682319254$\cdot 10^{-1}$ & 3.20611550947$\cdot 10^{-1}$ &
       2.01499388452207$\cdot 10^{7}$ & 6.414000669$\cdot 10^{6}$ \\

3500 & 3.13682319389$\cdot 10^{-1}$ & 3.20611550900$\cdot 10^{-1}$ &
       2.01499388452193$\cdot 10^{7}$ & 6.414000959$\cdot 10^{6}$ \\

3700 & 3.13682319454$\cdot 10^{-1}$ & 3.20611550894$\cdot 10^{-1}$ &
       2.01499388452230$\cdot 10^{7}$ & 6.414000992$\cdot 10^{6}$ \\

3840 & 3.13682319465$\cdot 10^{-1}$ & 3.20611550974$\cdot 10^{-1}$ &
       2.01499388452232$\cdot 10^{7}$ & 6.414000801$\cdot 10^{6}$ \\
       \hline\hline
$N({}^{4}$He) & $\langle \delta({\bf r}_{21}) \rangle$ & $\langle 
\delta({\bf r}_{31}) \rangle$ & $\langle \delta({\bf r}_{32}) \rangle$ & 
$\langle \delta({\bf r}_{321}) \rangle$ \\
       \hline
3300 & 3.13760535618$\cdot 10^{-1}$ & 3.20631791336$\cdot 10^{-1}$ &
       2.07001373516980$\cdot 10^{7}$ & 6.589975905$\cdot 10^{6}$ \\

3500 & 3.13760535756$\cdot 10^{-1}$ & 3.20631791288$\cdot 10^{-1}$ &
       2.07001373516964$\cdot 10^{7}$ & 6.589976203$\cdot 10^{6}$ \\

3700 & 3.13760535822$\cdot 10^{-1}$ & 3.20631791281$\cdot 10^{-1}$ &
       2.07001373516999$\cdot 10^{7}$ & 6.589976239$\cdot 10^{6}$ \\

3840 & 3.13760535832$\cdot 10^{-1}$ & 3.20631791364$\cdot 10^{-1}$ &
       2.07001373517002$\cdot 10^{7}$ & 6.589976038$\cdot 10^{6}$ \\
       \hline\hline
 \end{tabular}
 \end{center}
 \end{table}

\end{document}